\documentclass{ivs2e}       % for LaTeX2e class file
\usepackage{graphicx}
\usepackage{epstopdf}
\usepackage{wrapfig} 
\usepackage{color} 
\usepackage[font={footnotesize,up,singlespacing}]{caption}
\usepackage{multirow}

\Title{New Zealand VLBI Station, Warkworth}
\STitle{Warkworth Observatory}
\PublicationType{report}
\PublicationName{IVS 2014 Annual Report}
\IvsComponentName{Warkworth Observatory}
\NumberofAuthors{5}
\AuthorName{1}{Stuart Weston}
\AuthorName{2}{Tim Natusch}
\AuthorName{3}{Lewis Woodburn}
\AuthorName{4}{Ben Hart}
\AuthorName{5}{and Sergei Gulyaev}

\ContactAuthorReference{1}
\AuthorEmail{1}{stuart.weston@aut.ac.nz}
\AuthorTelephone{1}{+64 9 921 8689}
\AuthorPersonalWebPage{1}{n.a.}
\AuthorEmail{2}{tim.natusch@aut.ac.nz}
\AuthorTelephone{2}{+64 9 921 8689}
\AuthorPersonalWebPage{2}{n.a.}
\AuthorEmail{3}{lwoodbur@aut.ac.nz}
\AuthorTelephone{3}{+64 9 921 8689}
\AuthorPersonalWebPage{3}{n.a.}
\AuthorEmail{4}{ben.hart@aut.ac.nz}
\AuthorTelephone{4}{+64 9 921 8689}
\AuthorPersonalWebPage{4}{n.a.}
\AuthorEmail{5}{sergei.gulyaev@aut.ac.nz}
\AuthorTelephone{5}{+64 9 921 9541}
\AuthorPersonalWebPage{5}{n.a.}
\AuthorInstitutionReference{1}{1}
\AuthorInstitutionReference{2}{1}
\AuthorInstitutionReference{3}{1}
\AuthorInstitutionReference{4}{1}
\AuthorInstitutionReference{5}{1}

\NumberofInstitutions{1}
\InstitutionName{1}{Institute for Radio Astronomy and Space Research, Auckland University of Technology}
\InstitutionPostAddress{1}{Private Bag 92006, Auckland 1142}
\InstitutionCountry{1}{New Zealand}
\InstitutionFax{1}{n.a.}
\InstitutionWebPage{1}{http://www.irasr.ac.nz}

\begin{document}

%
% The command to make the title page from keywords in the header.
%
\MakeTitle                % not \maketitle

\setlength{\parindent}{0.5cm}
\setlength{\parskip}{0.1cm}
%%%%%%%%%%%%%%%%%%%%%%%%%%%%%%%%%%%%%%%%%%%%%%%%%%%%%%%%%%%%%%%
%
% Put your abstract text here.
%
\begin{abstract}
\noindent
The Warkworth Radio Astronomical Observatory is operated by the Institute for Radio Astronomy and Space Research (IRASR), AUT University, Auckland, New Zealand. 
Here we review the characteristics of the VLBI station facilities and report on a number of activities and technical developments in 2014.
\end{abstract}

\section{General Information}
% Section 1: General Information. Please provide general information about your component such as its location, its 
%sponsoring agency and what type of contribution you are making to IVS. If you have a photograph of your antenna, 
%correlator, analysis center, etc. you should include it.
%\vspace{-1cm}

\begin{figure}[htb!]
\begin{center}                   % center environment
  {
  \includegraphics[scale=0.25]{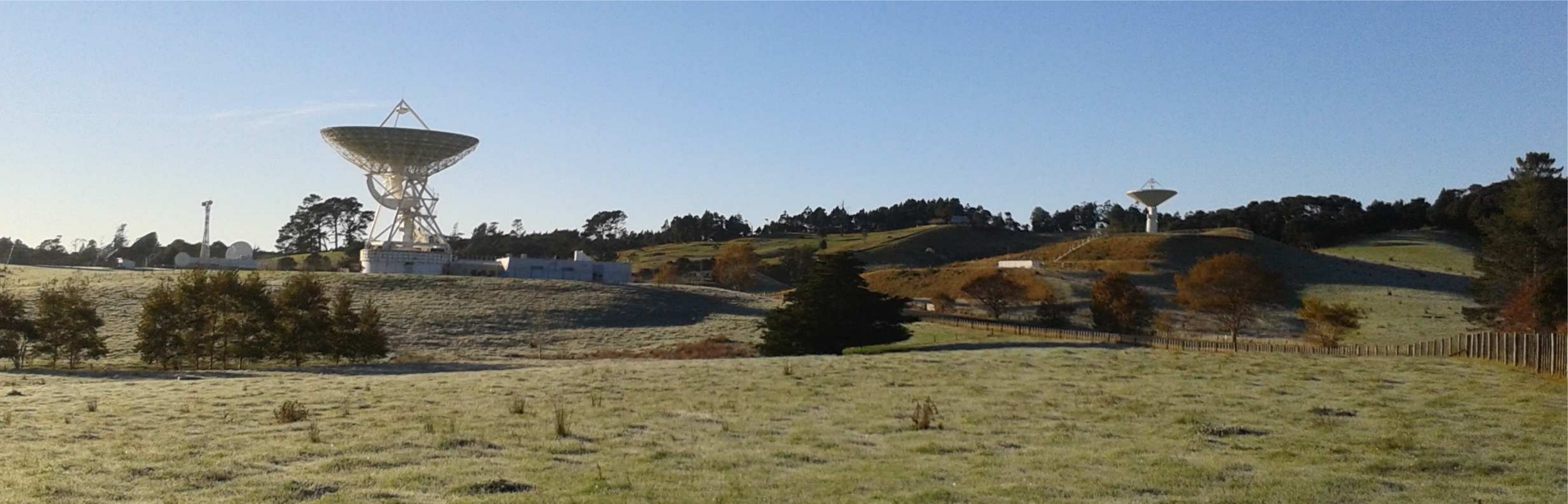}
  }
\caption{Photo of the two radio antennas at Warkworth on a frosty winters morning, on the left the 30-m and on the right the 12-m. In the background on the left hand side are the antennas belonging to Spark (formerly Telecom New Zealand). (Image courtesy of Stuart Weston)}
\label{fig:warkworth}
\end{center}
\end{figure}

The Warkworth Radio Astronomical Observatory  for which a panorama photo is shown in Figure \ref{fig:warkworth} is located some 60 km north of the city of Auckland, near the township of Warkworth. Specifications of the Warkworth 12-m and 30-m antennas are provided in Table \ref{t:wark}. The 12-m radio telescope is equipped with an S/X dual-band dual-circular polarization feed at the secondary focus and an L-band feed at the prime focus. Backend data digitizing is handled by a digital base band converter (DBBC) manufactured by the HAT-Lab, Catania, Italy. The 30-m radio telescope is currently equipped with a un-cooled C band dual-circular polarization receiver. The station frequency standard is a Symmetricom Active Hydrogen Maser MHM-2010 (75001-114). Mark 5B+ and Mark 5C data recorders are used for data storage and streaming of recorded data off site. The observatory network is directly connected to the national network provided by Research and Education Advanced Network New Zealand Ltd (REANNZ) via a 10~Gbps fibre link to the site \cite{r:karen}.

%\begin{wraptable}{l}{95mm}
% \begin{center}
\begin{table}[h]
   \centering
   \caption{Specifications of the Warkworth 12-m and 30-m antennas.}
%   \par\medskip\par
   \begin{tabular}{lll}
\hline
                                        & 12-m & 30-m    \\    
\hline
        Antenna type & Dual-shaped Cassegrain & wheel-and-track, Cassegrain\\
                               &                                          & beam-waveguide \\
        Manufacturer & Cobham/Patriot, USA & NEC, Japan \\
        Main dish Diam.      & 12.1 m  & 30.48 m \\
        Secondary refl. Diam. & 1.8 m & 2.715 m\\
        Focal length              & 4.538 m & \\
        Surface accuracy          & 0.35 mm & 1.2 mm\\
        Mount                     & alt-azimuth & alt-azimuth \\
        Azimuth axis range        & $90^\circ \pm 270^\circ$ &  $-179^\circ$ to $+354^\circ$ \\
        Elevation axis range      & $7.2^\circ$ to $88 ^\circ$ & $6.0^\circ$ to $90.1^\circ$\\
        Azimuth axis max speed    & $5^\circ$/s & $0.37^\circ$/s \\
        Elevation axis max speed  & $1^\circ$/s & $0.36^\circ$/s \\
        \hline
   \end{tabular}
   \label{t:wark}
\end{table}
% \end{center}
%\end{wraptable}
%        Frequency range           & 1.4---43~GHz \\
%        Main dish F/D ratio:      & 0.375 \\

\section{Component Description}
%Section 2: Component Description. Please provide a technical/scientific description of your component that is relevant to %your component type: for example, parameters of your network station's antenna, capabilities of your correlation center's %correlator, types of analysis solutions that you do as an analysis center, technology developments in progress at your %technology development center, etc.

\subsection{The 12-m antenna: Progress and issues}

In the beginning of 2014, a problem with slippage of the elevation axis encoder occurred. Initially, this issue could be dealt with by regenerating a pointing model on a fortnightly basis, but eventually it progressed to the point where a new model was required on an almost daily basis. It was determined that, as been the case at several other Patriot/Cobham 12-m antennas, the problem was caused by rotational slippage of the elevation axis pin connected to the encoder in the housing. As supplied by the manufacturer, this pin was intended to be locked to the main structure of the dish by an interface fit and convey rotational motion to the elevation encoder. A permanent solution that locks the pin to the structure of the dish through bolts drilled into the axis bearing pin appears to have been successful, we have not experienced any more drift and the elevation encoder offset has since been stable.

%A modifiction was made to the elevation bearing to stop the bronze surface sliding within the sleeve and causing the elevation encoder to drift in offset and thus invalidating the pointing model. This was particularly a problem during summer months and required re-pointing on almost a daily basis. Since this modification we have not experienced any more drift and the elevation encoder offset has stayed stable.

A new DBBC was received in April 2014 and was used to replace the original one purchased for the 12-m. This allowed the original DBBC to be returned to Bonn for repair and upgrade. Following its return, it has been installed as the digitiser for the 30-m antenna. Since the installation of the new DBBC on the 12-m we have seen a significant improvement in our SEFD figures. 
Previously they were of the order of 6000~Jy, now they are regularly $\sim 3800$~Jy, comparable with the AuScope 12-m antennas.

Having upgraded the Streamstor SDK for CONT14 and being able to address more than 1024 scans and greater than 16TB, we are in the process of upgrading our station diskpacks to 32TB.

\subsection{The 30-m antenna: Progress and issues}

\begin{figure}[hbp]   
\begin{minipage}[b]{0.45\linewidth} % A minipage that covers half the page, width-wise   
\centering 
 {
  \includegraphics[scale=0.6]{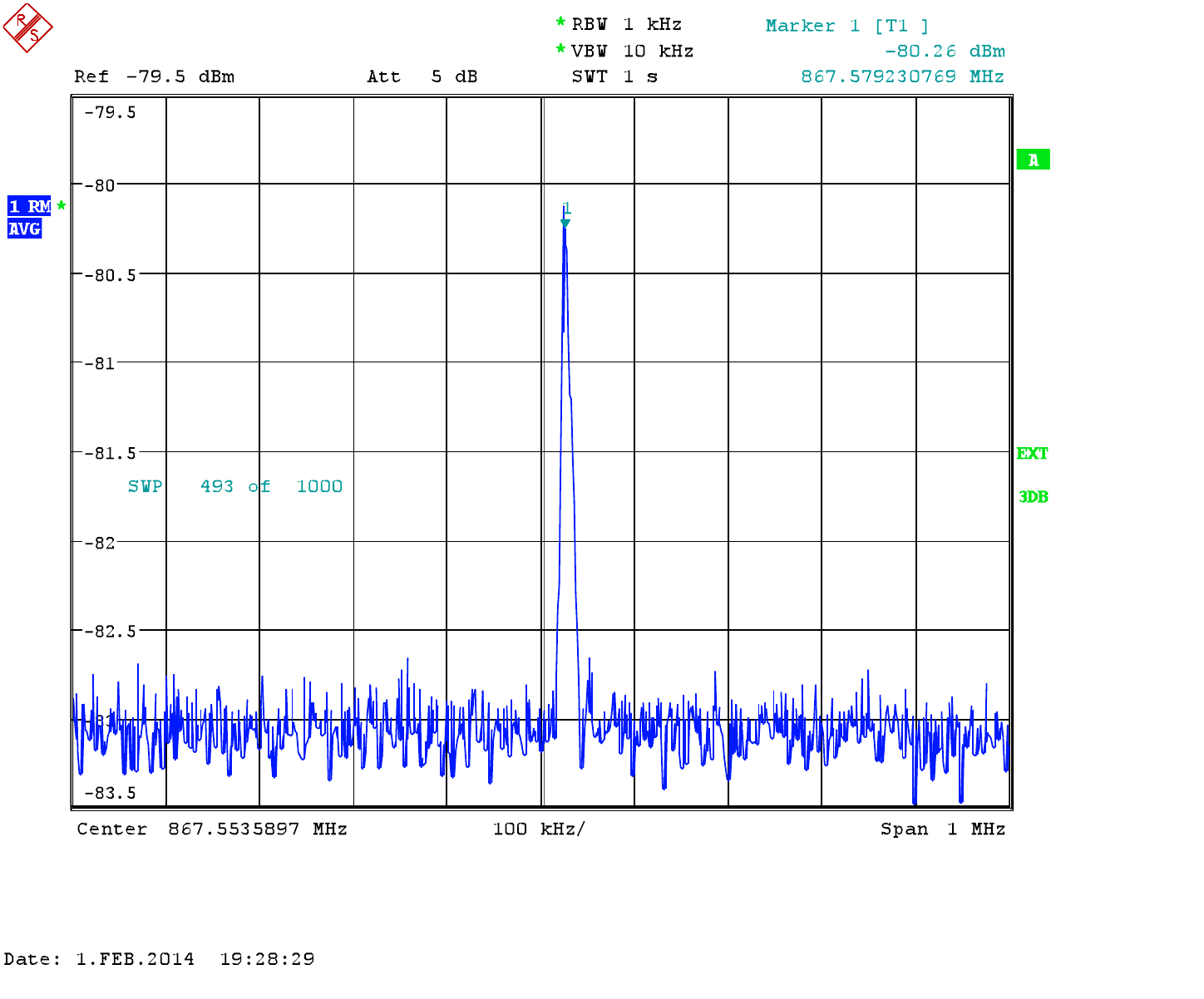}
  }
\caption{First light Methanol Maser Spectral line at 6.7 GHz. Credit: Tim Natusch}
\label{f:first_light_methanol}
\end{minipage}   
\mbox{\hspace{0.5cm}} % To get a little bit of space between the figures   
\begin{minipage}[b]{0.5\linewidth}   
\centering 
 {
  \includegraphics[scale=0.2]{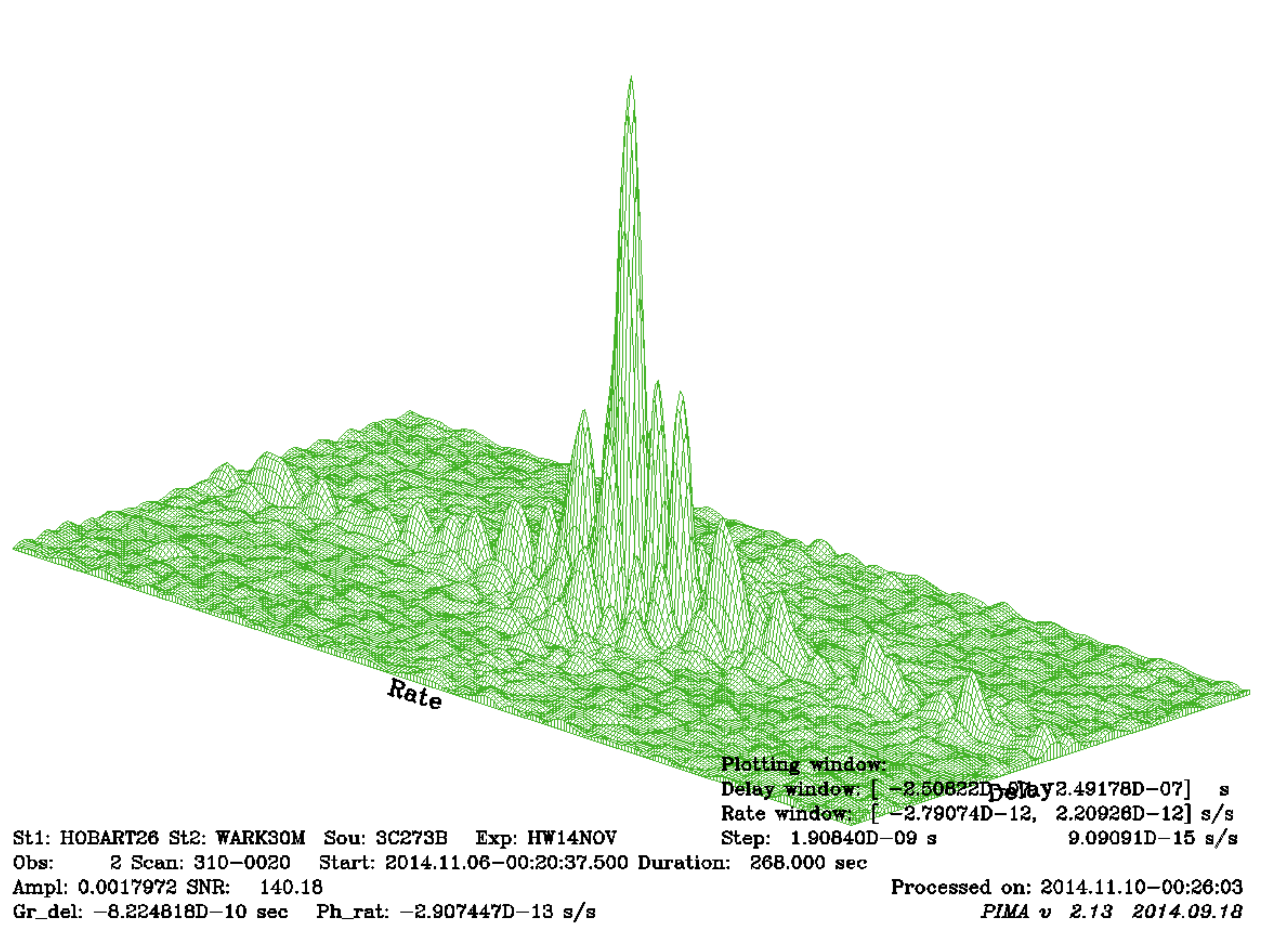}
  }
\caption{The first VLBI fringe between Warkworth 30-m and Hobart 26m at C-Band 6.7 GHz. Credit: Leonid Petrov}
\label{f:Ho_Wa_fringe}
\end{minipage}  
\end{figure}

By mid 2014 the conversion of the 30-m antenna had progressed to a stage where it was fully steerable and equipped with an uncooled 6~GHz receiver (donated by Jodrell Bank Observatory), and a First Light ceremony was held where a 6.7~GHz Methanol maser spectral line was received and displayed on a spectrum analyser Fig.~\ref{f:first_light_methanol}. More details of the conversion can be found in \cite{woodburn2014}.

With the return of the Observatory's original DBBC, installation of the Mk5B recorder and of a Symmetricom Universal Time and Frequency Distribution System to distribute signals from the observatory maser to the 30-m site, the antenna became capable of interferometry. The first VLBI fringe was detected on the baseline to the 26-m antenna at Hobart (Fig.~\ref{f:Ho_Wa_fringe}). We gratefully acknowledge the generous assistance of the University of Tasmania observatory in this accomplishment.

%The full digital backend was completed using the retured DBBC, one of our Mk5C's was down graded to a Mk5B+. Using this system in November for a VLBI test session between the UTas antennas as Hobart and Ceduna fringes were obtained at 6.7Ghz, see Figure \ref{f:Ho_Wa_fringe}. In 2015 it is planned to install a cooled C-Band receiver on the 30-m, we are also investigating wideband systems for both antennas.

\subsection{Warkworth Network}

In April 2014 the network link between Warkworth and the New Zealand research network (REANNZ) was upgraded to 10~Gbps. The international circuits from New Zealand provided by REANNZ are now 42~Gbps bi-directional to LA and 40~Gbps SXTransport ScienceWave (which REANNZ share with AARNet) that handles all the research and education routes. In October 2014 we installed 10~Gbps fibre network interface cards in our e-transfer Mk5. The bottle neck for us is now the receiving site, for example we can sustain 800~Mbps to the new data store in Perth run by iVEC, speeds to Bonn and USNO are of the order 100--200~Mbps.

\section{Current Status and Activities}

Year 2014 has seen a significant increase in the number of IVS sessions the 12-m has participated in, due in large part to the extra AUST sessions we are now able to participate in. A break down of session types (i.e. OHIG, CRDS, APSG, R and AUST) observed in 2014 are presented in Table \ref{t:ivs-wark}, showing our much higher utilisation/participation in IVS this year versus 2013.

\begin{table}[h]
   \centering
   \caption{The 12-m IVS 2014 Session Participation}
%   \par\medskip\par
   \begin{tabular}{l|cc}
        \hline
      \multirow{2}{*}{Experiment} & \multicolumn{2}{c}{Number of sessions}  \\
       & 2013 & 2014 \\
 \hline
        APSG & 2 & 2 \\
        OHIG & 3 & 6 \\
      R1 & 8 & 0 \\
       R4 & 8 & 11 \\
       CRDS & 6 & 6 \\
       CONT & 0 & 14 \\
       AUSTRAL & 6 & 57 \\
       AUST & 15 & 15 \\
        \hline
   \end{tabular}
   \label{t:ivs-wark}
\end{table}

The First geodetic observation with the Warkworth 30-m in C-Band with the Ceduna 30-m, Hobart 26 (University of Tasmania)  was conducted in December 2014. The schedule and analysis was undertaken by Leonid Petrov (NASA Goddard Space Flight Center), providing a preliminary solution for the 30-m antenna presented in Table \ref{t:ivs-wark-30m} (Petrov et al, 2015, in prep).

\begin{table}[h]
   \centering
   \caption{The 30-m preliminary solution components, Credit : Leonid Petrov}
%   \par\medskip\par
   \begin{tabular}{cccc}
        \hline
        Epoch & X, mm& Y, mm& Z, mm\\
    \hline
        2000.01.01-12:00:00 & -5115425635.06 & 477880304.86 & -3767042837.73  \\
       2014.12.11-12:59:43 & -5115425608.47 & 477880352.69 & -3767042708.48\\
        \hline
   \end{tabular}
   \label{t:ivs-wark-30m}
\end{table}

\noindent This observation will be repeated in 2015, it is hoped with additional stations in Japan (Tsukuba and Kashima) and South Africa (Hartebeesthoek) able to participate.

Foundations have been laid for the future hosting of a Gravimeter when in transit here in New Zealand from Antarctica. The site is just outside the 12-m antenna control facility so as to easily access utilities, and is also close to one of the LINZ (Land Information New Zealand) GNSS station hosted at the Warkworth Observatory. In March 2015 a site survey of the observatory will be undertaken with the assistance of LINZ. This will provide a useful check o fthe initial local tie survey of the 12-m antenna and GNSS station conducted at the end of 2012 \cite{gentle_2013}, and also provide a tie to the 30-m radio telescope for the first time.


\begin{thebibliography}{99} %99 sets the width of the widest expected
                            %entry label.  In this case, double-digit numbers
                            %will be created, and because 9 is as wide as any
                            %number, no label should be wider than 99.
\vspace{-0.3cm}
\bibitem{r:karen}
Weston, S., Natusch, T., Gulyaev, S., 
Radio Astronomy and e-VLBI using KAREN. {\it In Proceedings of the 17th Electronics New Zealand Conference}, 
2010.
Preprint arXiv:1011.0227.

\bibitem{woodburn2014}
Woodburn, L., Natusch, T., Weston, S., Thomasson, P., Godwin, M., Granet, C., Gulyaev, S.,
Conversion of a New Zealand 30 metre Telecommunications antenna into a Radio Telescope,
Publications of the Astronomical  Society of Australia,
Accepted, 
2015.

\bibitem{gentle_2013}
Gentle, P., Dawson, J., \& Woods, A.,
The 2012 Warkworth Observatory Local Tie Survey,
2013, 
 Tie Survey, Land Information New Zealand.

\vspace{-3cm}
\end{thebibliography}
\end{document}